\documentclass[preprintnumbers,amsmath,amssymb,prc,twocolumn]{revtex4-2}
\usepackage{graphicx}
\usepackage{amssymb}
\usepackage{amsmath}
\usepackage{ulem}
\usepackage{xcolor,etoolbox}
\usepackage{dcolumn}
\usepackage{bm}
\newcommand{\be}{\begin{equation}}
\newcommand{\ee}{\end{equation}}
\newcommand{\bea}{\begin{eqnarray}}
\newcommand{\eea}{\end{eqnarray}}

\newcommand{\comment}[1]{}
\renewcommand\sout{\bgroup \color{red} \ULdepth=-.5ex \ULset}

\begin{document}
\title{Neutron Skin Thickness Dependence of Astrophysical $S$-factor}

\author{T. Ghosh$^{1,2}$, Sangeeta$^3$, G. Saxena$^{4,5*}$, B. K. Agrawal$^{1,2}$, Ushasi Datta$^{1,2}$}
\affiliation{$^1$Saha Institute of Nuclear Physics, Kolkata-700064, India}
\affiliation{$^2$Homi Bhabha National Institute, Anushakti Nagar, Mumbai-400094, India}
\affiliation{$^3$Department of Applied Sciences, Chandigarh Engineering College-CGC, Landran-140307, India}
\affiliation{$^4$Department of Physics (H $\&$ S), Govt. Women Engineering College, Ajmer-305002, India}
\affiliation{$^5$Department of Physics, Faculty of Science, University of Zagreb, HR-10000 Zagreb, Croatia}
\begin{abstract}
\noindent\textbf{Background:} The density dependence of nuclear symmetry energy is crucial in determining several properties of finite nuclei to the neutron stars with mass $\sim$ 1.4 $M_\odot$. The values of neutron skin thickness, isovector giant dipole resonances energies and various nuclear reaction cross-sections in asymmetric nuclei have been utilized to determine the slope of symmetry energy ($L_0$) at the saturation density. Recent PREX-II and CREX measurements of neutron skin thickness in $^{208}$Pb and $^{48}$Ca nuclei yield very different values of $L_0$ which overlap marginally within 90\% confidence interval.\\
\textbf{Purpose:} Our objective is to demonstrate the role of symmetry energy on the sub-barrier fusion cross-section and the astrophysical $S$-factor for asymmetric nuclei.\\
\textbf{Method:} The nucleus–nucleus potentials are generated using the double-folding model (DFM) for three different nucleon-nucleon interactions. These DFM potentials are used for the calculation of sub-barrier fusion cross-section and the astrophysical $S$-factor. The nucleon densities required for DFM potentials are generated from different families of non-relativistic and relativistic mean-field models which correspond to a wide range of neutron skin thickness or $L_0$.\\
\textbf{Results:} We have calculated the sub-barrier fusion cross-section for several asymmetric nuclei involving O, Ca, Ni, and Sn isotopes. The results are presented for the barrier parameters, cross-section, and astrophysical $S$-factor for $^{54}$Ca+$^{54}$Ca and $^{124}$Sn+$^{124}$Sn as a function of neutron skin thickness.\\
\textbf{Conclusions:}  The cross-section for the neutron-rich nuclei show a strong dependence on the behavior of symmetry energy or the neutron skin thickness. The increase in skin thickness lowers the height of the barrier as well as its width which enhances the values of the $S$-factor by more than an order of magnitude. \\
\end{abstract}
\date{\today}

\maketitle

\section{Introduction}
The nuclear symmetry energy and its density dependence are instrumental in determining the properties of systems ranging from asymmetric nuclei to neutron stars and various other astrophysical phenomena \cite{lattimer2001,lattimer2023}. The neutron skin thickness in asymmetric nuclei is sensitive to the slope of symmetry energy ($L_{0}$) at the saturation density ($\rho_{0}$ $\sim$ 0.16 fm$^{-3}$) in a more or less model-independent manner \cite{brown2000,typel2001,yoshida2004,chen2005,centellas2009,reinhard2010,rocamaza2011,agrawal2012}. The value of surface symmetry energy to some extent depends upon $L_{0}$ that also governs the deformability of neutron-rich nuclei \cite{nikolov2011}. The equation of state for neutron star matter around the saturation density is predominantly governed by the density dependence of the nuclear symmetry energy. Consequently, the density content of symmetry energy is crucial in understanding a plethora of astrophysical phenomena such as the gravitational collapse of supernovae \cite{roberts2012,morozova2018}, neutron star crust’s thickness and thermal relaxation time in the neutron stars \cite{page2012}. The astrophysical observation by Neutron Star Interior Composition Explorer (NICER) \cite{nicer2016} and gravitational events from LIGO-Virgo \cite{ligo2016,ligo2017,ligo2021} have triggered many theoretical investigations to constrain the equation of state of neutron star matter.\par

The information on radius and tidal deformability derived from the astrophysical observations have been employed to constrain the slope and curvature parameters related to the density dependence of the symmetry energy \cite{malik2018,tsang2019,tsang2020,alam2016,carson2019,guven2020,malik2020,pradhan2022,xie2021}. The conclusions drawn from these studies are at variance as summarized in Ref. \cite{kunjipurayil2022}. It is shown very recently \cite{patra2023} that the correlations of radius and tidal deformability of neutron stars with symmetry energy parameters are quite sensitive to the choice of distributions of these parameters and various other factors. These correlations almost vanish if the various symmetry energy parameters are considered to be independent of each other.\par

Recently, neutron skin thickness for $^{208}$Pb and $^{48}$Ca nuclei have been determined through parity-violating electron scattering experiments
PREX-II \cite{prex2021} and CREX \cite{crex2022}, respectively. The measured values of neutron skin thickness ($\Delta$r$_{np}$) are 0.283$\pm$0.071 fm  for $^{208}$Pb and 0.121$\pm$0.026(exp)$\pm$0.024(model) fm for $^{48}$Ca nucleus. The $\Delta$r$_{np}$ ($^{208}$Pb) leads to $L_{0}$=106$\pm$37 MeV for a specific class of relativistic mean-field energy density functionals \cite{reed2021}. The values of $L_{0}$= 76$\sim$165 MeV from $\Delta$r$_{np}$ ($^{208}$Pb) and  $L_{0}$= 0$\sim$51 MeV from $\Delta$r$_{np}$ ($^{48}$Ca) have been deduced by using 207 non-relativistic and relativistic mean-field models \cite{tagami2022}.
The combined analysis of PREX-II and CREX data results in $L_{0}$=15.3$^{+46.8}_{-41.5}$ MeV with 90$\%$ of confidence level which yields $\Delta$r$_{np}$ ($^{208}$Pb)= 0.139$^{+0.070}_{-0.060}$ fm and $\Delta$r$_{np}$ ($^{48}$Ca) = 0.140$^{+0.035}_{-0.032}$ fm \cite{Zhang2022}. Most of the mean-filed models that are consistent with nuclear masses and charge radii throughout the nuclear chart as well as the available data of dipole polarizabilities do not simultaneously explain the measured values of $\Delta$r$_{np}$ for $^{208}$Pb and $^{48}$Ca nuclei \cite{reinhard2022}.\par

Several alternative ways to investigate the density dependence of the symmetry energy or the neutron skin thickness have been proposed. One such proposal
is to measure $\Delta$r$_{np}$ through the coherent elastic neutrino-nucleus scattering with a neutrino flux from a nearby core-collapse supernova in our galaxy \cite{huang2022}. The sensitivity of symmetry energy to nuclear reactions has also been lately investigated \cite{reinhard2016,tagami2021,colomer2022,wakasa2022}. The fusion cross-section of $^{48}$Ca+$^{48}$Ca near the barrier manifests the dependence on the symmetry energy slope parameter. The proton-nucleus scattering is also found to be dependent on the values of $\Delta$r$_{np}$ or the $L_{0}$ \cite{wakasa2022}.

In the present paper, we study the sensitivity of density dependence of the symmetry energy to the sub-barrier fusion cross-sections and the resulting astrophysical $S$-factor for a few asymmetric nuclei. The density profiles which are one of the key inputs to the double-folding model potentials and $S$-factor calculations are obtained from several parameterizations of non-relativistic and relativistic mean-field models corresponding to different values of neutron skin thickness or $L_{0}$. Our results are quite sensitive to the values of neutron skin thickness.

\section{Methodology}
Low-energy heavy-ion fusion reactions are governed by quantum tunneling through the Coulomb barrier formed by the combination of repulsive Coulomb force and attractive nuclear interaction. From the partial wave analysis of formal nuclear reaction theory \cite{Blatt1952} results the following formula of the cross-section for two nuclei undergoing nuclear reaction
\begin{equation}
    \sigma (E) = \dfrac{\pi}{k^2} \sum_{l=0}^{\infty} (2l + 1) T_{l}(E)
    \label{eq:sigma}
\end{equation}
where, $k=\frac{\sqrt{2\mu E}}{\hbar}$ , $\mu$ being the reduced mass of the interacting nuclei and the transmission coefficient $T_{l}(E)$ for $l^{th}$ partial wave is given by
\begin{equation}
    T_{l}(E) = exp (-\dfrac{2}{\hbar}\int_{r_{1}}^{r_{2}} \sqrt{2\mu [V_{eff}(r) - E]} dr)
\end{equation}
within WKB approximation, where $r_1$ and $r_2$ are classical turning points, $E$ is the energy in the centre of mass frame and $V_{eff}(r)$ is the effective barrier potential expressed as
\begin{equation}
    V_{eff}(r) = V_{n}(r) + V_{c}(r) + \dfrac{l(l+1)\hbar^2}{2\mu r^2}
\end{equation}
The three terms in the above equation account for the potential energy arising out of nuclear, coulomb and centrifugal force, respectively.\par
There are several ways to estimate the nuclear potential $V_{n}(r)$. One standard method is to fold a nucleon-nucleon interaction with the nucleonic density distribution of projectile and target \cite{Satchler1979}.
In our calculations, the double-folding model (DFM) is used to generate the nucleus–nucleus potential which reads as,
\begin{equation}
    V_n(r)=\int \int V_{NN} (\lvert \textbf{R} - \textbf{r}_t + \textbf{r}_p \rvert )\rho_p(\textbf{r}_p) \rho_t(\textbf{r}_t) d\textbf{r}_p d\textbf{r}_t
    \label{Eq:DFM}
\end{equation}
here $\textbf{r}_p$ and $\textbf{r}_t$ are the radius vectors of two interacting points of the target and projectile nuclei, respectively; $\textbf{R}$ denotes the vector joining their centers of mass; $\rho_p(\textbf{r}_p)$ and $\rho_t(\textbf{r}_t)$ stand for the target and projectile nuclear matter densities \cite{Gontchar2010,Gontchar2021,Chamon2007,Chamon2021}.
DFM potentials are calculated using the code for the double-folding interaction potential of two spherical nuclei (DFMSPH22) \cite{Gontchar2010,Gontchar2021} employing M3Y-Paris parameterizations with (DDM3Y1) and without density dependence of the exchange part of the NN-forces, $V_{NN}$. São Paulo Potential version 2 (SPP2) \cite{Chamon2007,Chamon2021} computer code is also used for the comparison. The nucleonic densities entering DFM are generated from Skyrme-Hartree-Fock-Bogoliubov models \cite{Vautherin1972, Vautherin1973,Dutra2012} and various RMF models \cite{Lalazissis1997,Todd-Rutel2005,Lalazissis2005} corresponding to different values of $L_0$. These DFM potentials are used in a special version of single-channel CCFULL code \cite{CCFULLwebpage,Hagino1999} to calculate fusion cross-section ($\sigma$(E)). \par


The cross-section given by Eq. \ref{eq:sigma} decreases exponentially at deep sub-barrier energies of astrophysical relevance due to the larger barrier height and width encountered. The expression of the cross-section can be decomposed into the term that depends strongly on energy and another one that varies weakly with energy as,
\begin{equation}
   \sigma(E) =E^{-1} exp(-2\pi \eta) S(E)
    \label{Eq:s-factor}
\end{equation}
S(E), called the astrophysical $S$-factor, varies slowly with energy and it includes various nuclear structure effects. Here, E is the center of mass energy of the reactants, $\eta=\frac{Z_{1}Z_{2}e^{2}}{\hbar \nu}$ is the Sommmerfeld parameter and $\nu=\sqrt{\frac{2E}{\mu}}$ denotes the relative velocity of the reactants at large separations.

\section{Sub-barrier fusion cross-section and astrophysical $S$-factor} The density dependence of the symmetry energy significantly influences the neutron-skin thickness in asymmetric nuclei which is not very well determined to date. We have studied the dependence of cross-section for the sub-barrier fusion of asymmetric nuclei on the neutron-skin thickness. One of the important inputs to such calculations is the distributions of nucleons inside the nuclei which are employed to construct DFM potentials using suitable nucleon-nucleon interactions. The density profiles of the protons and neutrons for a given nucleus are obtained using non-relativistic and relativistic mean-field models.

\begin{table}[!htb]
\caption{\label{tab1} The energy per nucleon $\epsilon_0$, incompressibility co-efficient $K_0$, effective nucleon mass $m^*_{0}$, symmetry energy co-efficient $J_0$, slope $L_0$ and curvature $K_{sym,0}$ of the symmetry energy, evaluated for the nuclear matter at the saturation density $\rho_{0}$ using the SLy4 and SkO effective interactions \cite{Chabanat1998,Reinhard1999}. The values of $\rho_{0}$ are in fm$^{-3}$, m* are in the units of the bare mass of nucleon, and the remaining parameters are in MeV.}
\centering
\resizebox{0.5\textwidth}{!}{%
\begin{tabular}{c | c @{\hskip 0.1in} c @{\hskip 0.1in} c @{\hskip 0.1in}c @{\hskip 0.1in}c@{\hskip 0.1in} c@{\hskip 0.1in} c  }
\hline
\multicolumn{1}{c|}{Model}&
\multicolumn{1}{c}{$\rho_{0}$}&
\multicolumn{1}{c}{$\epsilon_0$}&
\multicolumn{1}{c}{$K_{0}$}&
\multicolumn{1}{c}{$m^*_{0}$}&
\multicolumn{1}{c}{$J_0$}&
\multicolumn{1}{c}{$L_0$}&
\multicolumn{1}{c}{$K_{sym,0}$}\\

\hline
SLy4 & 0.16 & -15.97 & 521.53 & 0.69 &	229.91 & 45.94 & -119.73 	 \\
SkO & 0.16 & -15.84 & 131.13  & 0.90 &	223.34 &	79.14 &-43.17 \\
\hline
\end{tabular}}
\end{table}

\begin{figure}[!htb]
\begin{center}
\includegraphics[width=0.49\textwidth]{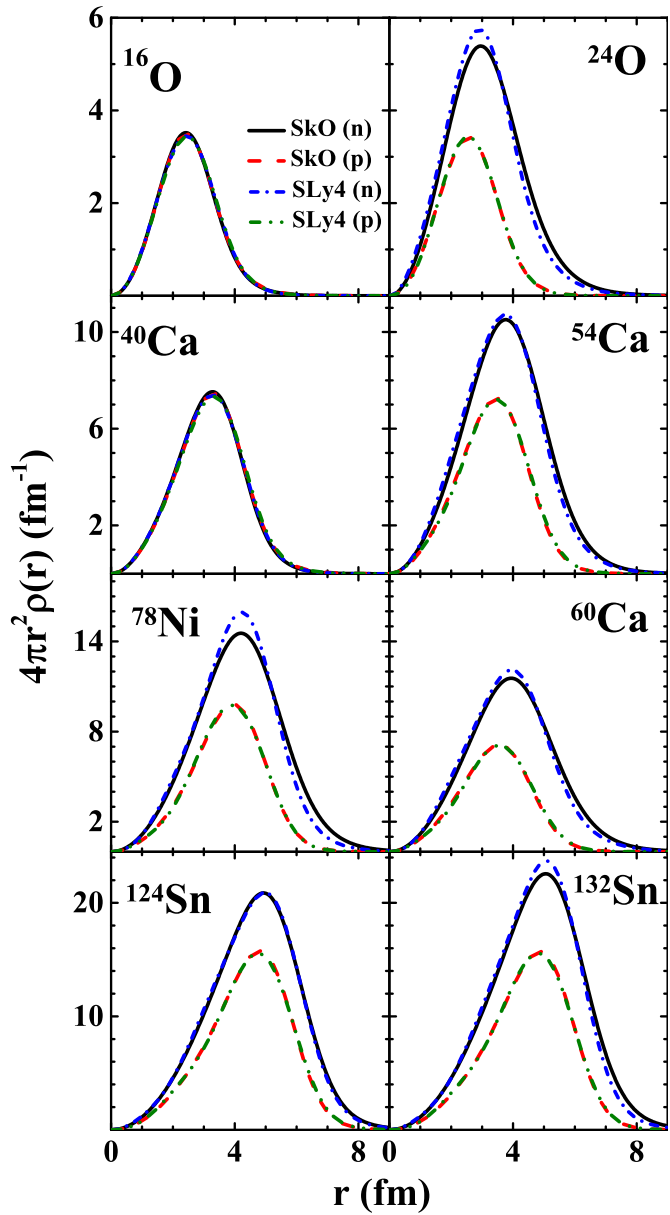}
\end{center}
\vspace{-0.5cm}
\caption{Radial density distributions for neutrons(n) and protons(p) for several nuclei, obtained from SLy4 and SkO Skyrme effective interactions.}\label{den}
\end{figure}

We have used non-relativistic mean-field models derived from the Skyrme type effective interactions \cite{Vautherin1972, Vautherin1973}. Two different variants of relativistic mean-field models employed are: (i) includes contributions from the non-linear self- and mixed-interactions of the mesons with constant coupling strengths \cite{Lalazissis1997,Todd-Rutel2005} and (ii) the nonlinearity of mesons fields are incorporated through the density-dependent coupling constants \cite{Lalazissis2005}. The double-folding model potentials are obtained using the density distributions of nucleons in the nucleus from the mean-field model together with the M3Y-Paris potential without (PDD0) and with (PDD1) density dependence as well as São Paulo potential (SPP2), as briefly outlined in the previous section. \par

We first present our results in detail for SLy4 and SkO Skyrme effective interactions \cite{Chabanat1998, Reinhard1999} which mainly differ in the behavior of the symmetry energy. In Table \ref{tab1}, we list the various properties that characterize the uniform nuclear matter at the saturation density. The energy per nucleon $\epsilon_0$, incompressibility co-efficient $K_0$, and effective nucleon mass $m^*_{0}$ describe the symmetric nuclear matter, whereas, the density dependence of the symmetry energy is described by symmetry energy co-efficient $J_0$, slope $L_0$, and curvature $K_{sym,0}$. The values of $L_0$ and $K_{sym,0}$ are significantly different for the Skyrme effective interactions considered. In Table \ref{tab:skin}, we present the Skyrme Hartree-Fock results for a few nuclei of our interest. Both the Skyrme effective interactions considered yield very similar values for the protons rms radii, but, a different value for neutron skin thickness $\Delta$r$_{np}$ = r$_n$-r$_p$ with r$_n$ and r$_p$ being the rms radii for neutron and proton, respectively, obtained from the point density distributions. The values of $\Delta$r$_{np}$ are larger for SkO interaction due to the larger value of $L_0$ \cite{centellas2009}. The difference between neutron skin thickness from SkO and SLy4 interactions is also listed in the last column. The choice of SkO and SLy4 Skyrme interactions will enable us to examine the dependence of sub-barrier fusion cross-section on $\Delta$r$_{np}$ or $L_0$. \par

\begin{table}[!htb]
\caption{\label{tab:skin} The values of binding energy per nucleon (BE/A), rms radius for proton (r$_{p}$), and neutron skin thickness ($\Delta$r$_{np}$) obtained for SLy4 and SkO Skyrme effective interactions. The BE/A is in MeV whereas r$_{p}$, $\Delta$r$_{np}$, and the difference $\Delta$r$_{np}^{diff}$ = $\Delta$r$_{np}$(SkO)-$\Delta$r$_{np}$(SLy4) are in the units of fm.}
\centering
\resizebox{0.5\textwidth}{!}{%
\begin{tabular}{c | r@{\hskip 0.1in} c @{\hskip 0.1in}c @{\hskip 0.1in}| r@{\hskip 0.1in} c @{\hskip 0.1in}c @{\hskip 0.1in}|c}
\hline
\multicolumn{1}{c|}{Nucleus}&
\multicolumn{3}{c|}{SLy4}&
\multicolumn{3}{c}{SkO}&
\multicolumn{1}{|c}{$\Delta$r$_{np}^{diff}$}\\
\cline{2-7}
\multicolumn{1}{c|}{}&
\multicolumn{1}{c}{BE/A}&
\multicolumn{1}{c}{r$_{p}$}&
\multicolumn{1}{c|}{$\Delta$r$_{np}$}&
\multicolumn{1}{c}{BE/A}&
\multicolumn{1}{c}{r$_{p}$}&
\multicolumn{1}{c}{$\Delta$r$_{np}$}&
\multicolumn{1}{|c}{}\\

\hline
$^{16}$O &8.029 &	2.702&	-0.023& 7.406 &	2.694  &-0.021&0.002 \\
$^{24}$O & 7.208 &	2.762 &	0.479  & 6.950 & 2.760  & 0.628&0.149 \\
$^{40}$Ca &8.606 &	3.419&	-0.047& 8.397 &	3.401&	-0.041&0.006\\
$^{54}$Ca & 8.311 & 3.534 & 0.350 & 8.267 & 3.509 & 0.466&0.116\\
$^{60}$Ca &7.750 &	3.615&	0.483& 7.818 &	3.584&	0.632&0.149\\
$^{78}$Ni &8.253 &	3.925 &	0.310 & 8.254 &	3.929 &	0.451&0.141 \\
$^{124}$Sn & 8.472 & 4.626 & 0.180 & 8.471 & 4.614 & 0.226&0.046\\
$^{132}$Sn &8.358 &	4.670&	0.222& 8.320 &	4.670&	0.320&0.098\\
\hline
\end{tabular}}
\end{table}

In Fig. \ref{den}, we display the radial variation of density for several nuclei. The SkO Skyrme interaction yields neutron densities quite different in the tail region than those for the SLy4 interaction. The behavior of the sub-barrier fusion cross-sections is sensitive to these differences in the density profiles which may be more pronounced in the case of highly asymmetric nuclei.
\begin{figure}[!htb]
\begin{center}
\includegraphics[width=0.49\textwidth]{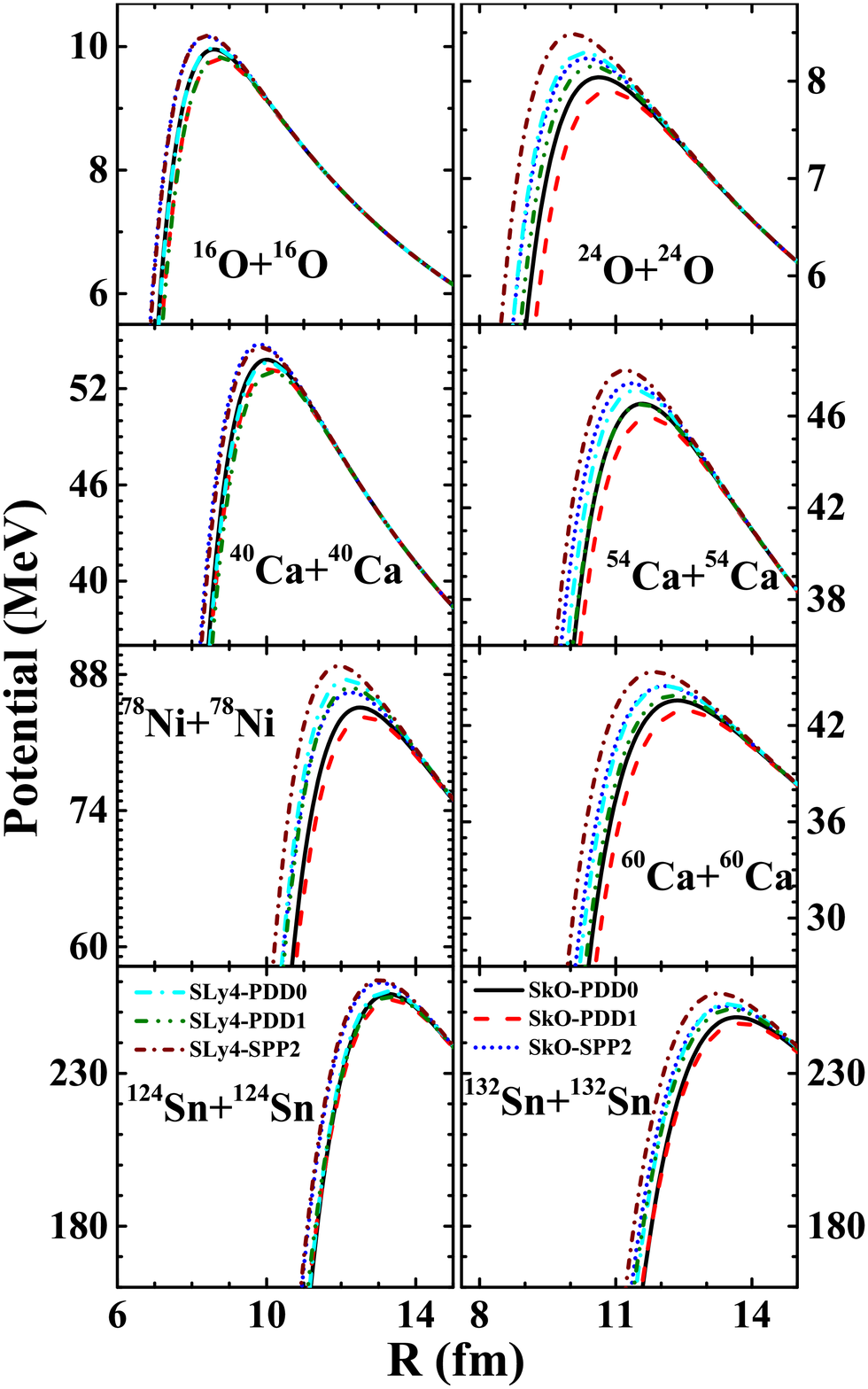}
\end{center}
\vspace{-0.5cm}
\caption{Potentials obtained by folding the density profiles of Fig. \ref{den} with the different nucleon-nucleon interactions as a function of the distance between the center of masses of two nuclei. The curves level as PDD0, PDD1, and SPP2 correspond to the folding potentials M3Y-Paris without density dependence, M3Y-Paris with density dependence, and São Paulo potential version 2, respectively.}\label{pot}
\end{figure}

In Fig. \ref{pot}, we show the DFM potentials as a function of the distance of the center of masses of two nuclei. These potentials are obtained by folding the nuclear densities with the PDD0, PDD1, and SPP2 nucleon-nucleon interactions \cite{Gontchar2010,Gontchar2021,Chamon2007,Chamon2021}. The qualitative behavior for all the DFM potentials is similar. The maximum barrier heights and the widths at a given energy are larger for SPP2 followed by PDD0 and PDD1 potentials. The SkO interaction associated with larger skin thickness yields smaller barrier heights and widths.

\begin{table}[!htb]
\caption{\label{tab:barrier} The maximum barrier height ($V_{B}$) and width (W) obtained by DFM potentials M3Y-Paris without density dependence (PDD0), M3Y-Paris with density dependence (PDD1), and São Paulo potential version 2 (SPP2) \cite{Gontchar2010,Gontchar2021,Chamon2007,Chamon2021} together with the density profiles from SLy4 and SkO Skyrme effective interactions. The width of the barrier for a given nucleus is calculated at the same energy corresponding to the 0.7 times barrier height for the SLy4 interaction. The values of $V_{B}$ are in MeV whereas W and neutron skin thickness $\Delta$r$_{np}$ are in fm.}
\centering
\resizebox{0.5\textwidth}{!}{%
\begin{tabular}{c | c | r@{\hskip 0.1in} c @{\hskip 0.1in}c @{\hskip 0.1in} |r@{\hskip 0.1in} c @{\hskip 0.1in}c @{\hskip 0.1in}  }
\hline
\multicolumn{1}{c|}{Nucleus}&
\multicolumn{1}{c|}{Pot.}&
\multicolumn{3}{c|}{SLy4}&
\multicolumn{3}{c}{SkO}\\
\cline{3-8}
\multicolumn{1}{c|}{}&
\multicolumn{1}{c|}{}&
\multicolumn{1}{c}{$V_{B}$}&
\multicolumn{1}{c}{W }&
\multicolumn{1}{c|}{$\Delta$r$_{np}$}&
\multicolumn{1}{c}{$V_{B}$}&
\multicolumn{1}{c}{W }&
\multicolumn{1}{c}{$\Delta$r$_{np}$}\\

\hline
$^{24}$O &PDD0 & 8.293 & 7.375 & 0.479 & 8.041 & 7.080 &0.628 \\
  & PDD1 & 8.152 & 7.204 & 0.479 & 7.894 & 6.876 &0.628 \\
 & SPP2 & 8.484 & 7.638 & 0.479 & 8.234 & 7.372 &0.628 \\
$^{54}$Ca & PDD0 & 47.115 & 7.915 & 0.350 & 46.539 & 7.786 & 0.466\\
   & PDD1 & 46.514 & 7.806 & 0.350 & 45.914 & 7.661 & 0.466\\
    & SPP2 & 47.997 & 8.165 & 0.350 & 47.425 & 8.049 & 0.466\\
$^{60}$Ca &PDD0 & 44.461 & 8.409 & 0.483& 43.556 & 8.197 &0.632\\
 & PDD1 & 43.852 & 8.274 &  0.483& 42.910 & 8.033 &0.632\\
   & SPP2 & 45.337 & 8.679 & 0.483 & 44.451 & 8.492 & 0.632\\
$^{78}$Ni &PDD0 & 87.467 & 8.158 &  0.310& 84.605 & 7.873 &0.451 \\
 & PDD1 & 86.613 & 8.104 & 0.310 & 83.582 & 7.771 &0.451\\
    & SPP2 & 88.862 & 8.389 & 0.310 & 86.139 & 8.142 &0.451 \\
$^{124}$Sn & PDD0 & 256.846 & 8.851 & 0.180 & 255.928 & 8.822 &0.226 \\
   & PDD1 & 254.974 & 8.881 & 0.180 & 253.994 & 8.850 & 0.226\\
    & SPP2 & 260.463 & 9.087 & 0.180 & 259.623 & 9.063 &0.226 \\
$^{132}$Sn &PDD0 &	252.675 & 8.905 &  0.222&248.349 & 8.753 &0.320\\
 & PDD1 & 250.950 & 8.940 &  0.222& 246.314 & 8.758 & 0.320\\
   & SPP2 & 256.108 & 9.134 & 0.222 & 252.027 & 9.003 & 0.320\\
\hline
\end{tabular}}
\end{table}

In Table \ref{tab:barrier}, we list the values of maximum barrier height, width at a fixed energy, and neutron skin thickness for all the asymmetric nuclei considered. The results are presented for both the Skyrme effective interactions and with all the three DFM potentials. The effects of neutron skin thickness on various barrier parameters are evident. These effects become stronger with increasing proton numbers. For instance, the maximum change in the barrier parameters with the change in the skin thickness from SLy4 to SkO Skyrme interaction is seen for $^{132}$Sn, though, it is less asymmetric than the $^{60}$Ca. The reductions in height is about 4 MeV and in the width is about 0.15 fm with the increase in neutron skin thickness by 0.1 fm for the case of $^{132}$Sn. Whereas, for $^{60}$Ca the reduction in barrier height is 1 MeV and the width is 0.1 fm with an increase in neutron skin thickness by 0.15 fm. The area of the barrier is determined by its height and width that governs the values of cross-section exponentially. Thus, small changes in the barrier parameters could significantly affect the cross-section and astrophysical $S$-factor. \par

\begin{figure}[!htb]
\begin{center}
\includegraphics[width=0.49\textwidth]{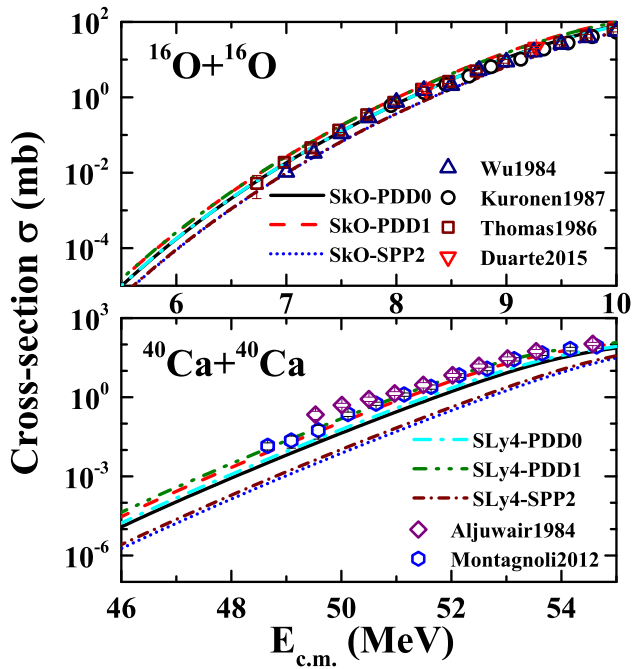}
\end{center}
\vspace{-0.5cm}
\caption{Cross-sections for the sub-barrier fusion of symmetric systems $^{16}$O+$^{16}$O and $^{40}$Ca+$^{40}$Ca as a function of energy in the center of mass frame (E$_{cm}$) calculated for different DFM potentials and density profiles. The experimental data are taken from Refs. \cite{Wu1984,Aljuwair1984,Thomas1986,Kuronen1987,Montagnoli2012,Duarte2015}.}\label{cs-2p}
\end{figure}

\begin{figure}[!htb]
\begin{center}
\includegraphics[width=0.49\textwidth]{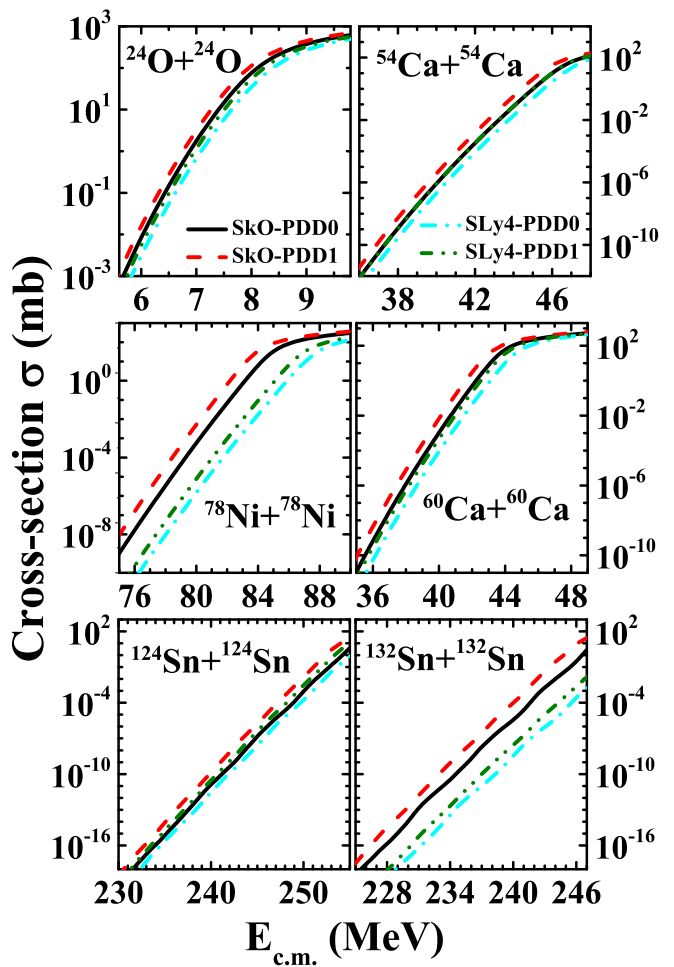}
\end{center}
\vspace{-0.5cm}
\caption{Same as Fig. \ref{cs-2p}, but for several asymmetric systems where the cross-sections are obtained with M3Y-Paris without density-dependent (PDD0), and with density-dependent (PDD1) potentials.}\label{cs-6p}
\end{figure}

Before embarking on our results for the fusion of asymmetric systems, we present the variation of cross-sections ($\sigma_f$) as a function of energy in the center of mass frame for the symmetric systems such as $^{16}$O+$^{16}$O and $^{40}$Ca+$^{40}$Ca in Fig. \ref{cs-2p}. We have also shown the corresponding experimental data \cite{Wu1984,Aljuwair1984,Thomas1986,Kuronen1987,Montagnoli2012,Duarte2015} for comparison. From a close inspection of the figure, it is found that SPP2 underestimates the cross-sections and hereafter we will consider the results only for PDD0 and PDD1 potentials. In Fig. \ref{cs-6p}, we show the values of sub-barrier fusion cross-sections for asymmetric systems, $^{24}$O+$^{24}$O, $^{54}$Ca+$^{54}$Ca, $^{60}$Ca+$^{60}$Ca, $^{78}$Ni+$^{78}$Ni, $^{124}$Sn+$^{124}$Sn, and $^{132}$Sn+$^{132}$Sn. Unlike the case of symmetric nuclei, the degeneracy in the cross-section obtained by using SLy4 and SkO effective interactions disappears. The SkO interaction which yields larger skin thickness leads to a larger cross-section due to the reduction in barrier height and width. The cross-section for $^{24}$O+$^{24}$O is enhanced by three to four times due to increase in skin thickness by 0.15 fm for SkO in comparison with those of SLy4 interaction (see the last column of Table \ref{tab:skin}). For $^{132}$Sn+$^{132}$Sn case, the cross-section enhances by two to three orders of magnitude due to increase in neutron skin thickness by 0.10 fm. These trends may be indicative of the fact that the Coulomb potentials for the nuclei with larger proton numbers are more sensitive to the change in nuclear size or the neutron skin thickness, since, proton radii are more or less fixed. \par

\begin{figure}[!htb]
\begin{center}
\includegraphics[width=0.49\textwidth]{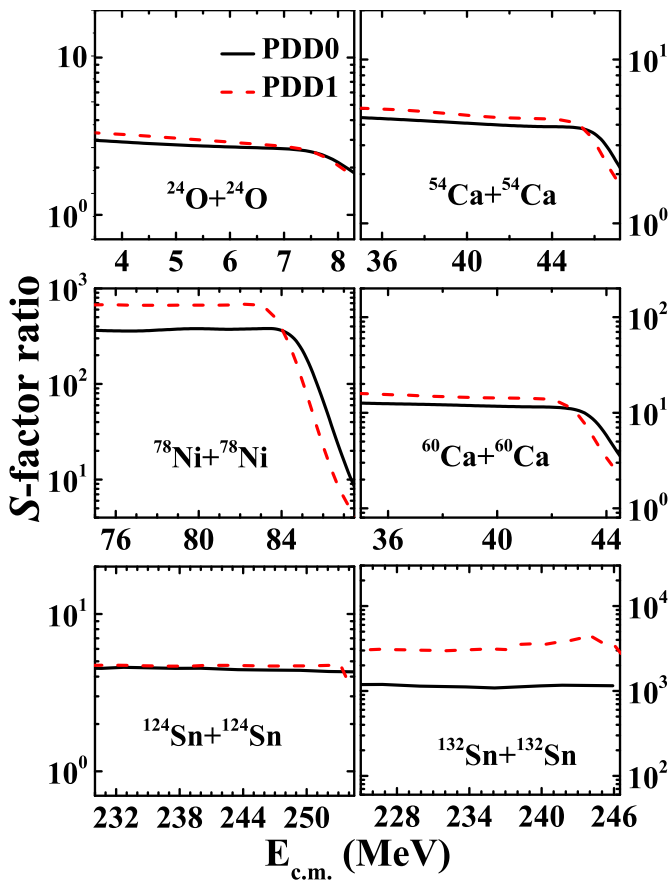}
\end{center}
\vspace{-0.5cm}
\caption{Ratio of $S$-factor obtained by using nucleonic density distribution for the SkO to the Sly4 Skyrme forces with different DFM potentials as indicated in the figure.}\label{s-factor-ratio}
\end{figure}

We calculate the astrophysical $S$-factor, which is directly related to the cross-section (Eq. \ref{Eq:s-factor}), for the fusion of various asymmetric nuclei considered. In Fig. \ref{s-factor-ratio}, we plot the ratio of the $S$-factor obtained for the SkO interaction to those for the SLy4 interaction. These results are obtained with PDD0 and PDD1 folding potentials. For the cases of $^{24}$O+$^{24}$O, $^{60}$Ca+$^{60}$Ca and $^{78}$Ni+$^{78}$Ni the increase in neutron skin thickness for SkO with respect to the SLy4 force is $\sim$0.15 fm. The rapid increase in the values of the $S$-factor in these nuclei for SkO force clearly suggests its sensitivity to the neutron skin thickness, which grows stronger with the increase in proton number. The $S$-factor increases by one order of magnitude for $^{54}$Ca+$^{54}$Ca with the increase in skin thickness by 0.1 fm and a similar increase is observed for $^{124}$Sn+$^{124}$Sn with the increase in skin thickness only by 0.05 fm (see the Table \ref{tab:skin}).\par

\begin{figure}[!htb]
\begin{center}
\includegraphics[width=0.49\textwidth]{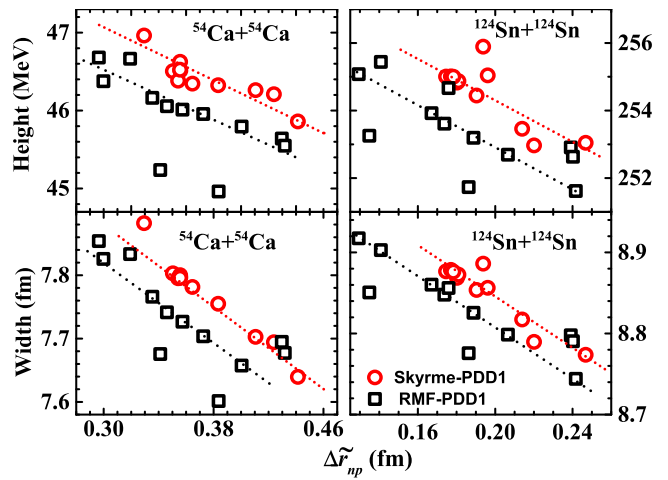}
\end{center}
\vspace{-0.5cm}
\caption{Values of maximum barrier height and the width as a function of neutron skin thickness $\Delta$$\Tilde{r}_{np}$ for different Skyrme effective interactions (red circles) and relativistic models (black squares). The values of $\Delta$$\Tilde{r}$$_{np}$ are calculated with respect to the proton radii obtained for SLy4 effective interaction. The width of the barrier is calculated at 32.56 MeV for $^{54}$Ca+$^{54}$Ca and at 178.48 MeV for $^{124}$Sn+$^{124}$Sn which corresponds to 0.7 times the barrier height for the SLy4 effective interaction. All the red circles and black squares are connected separately to guide the eyes.} \label{bar}
\end{figure}

\begin{figure}[!htb]
\begin{center}
\includegraphics[width=0.49\textwidth]{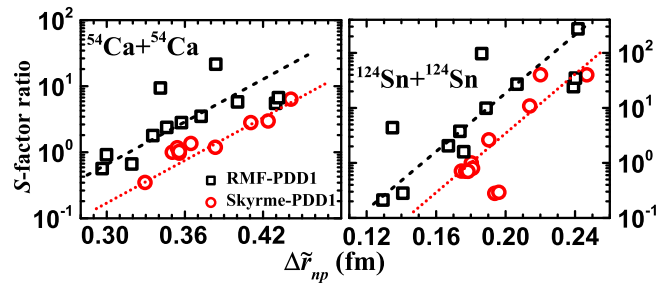}
\end{center}
\vspace{-0.5cm}
\caption{Similar to Fig. \ref{bar}, but, for the $S$-factor ratio for a given model with respect to the $S$-factor obtained for the SLy4.}\label{cs-s}
\end{figure}
We also calculate the cross-section and $S$-factor by varying the neutron skin thickness for $^{54}$Ca+$^{54}$Ca and $^{124}$Sn+$^{124}$Sn, which might be within the experimental reach. We fix the proton distribution to the ones obtained for SLy4 interaction and the distribution of neutrons is obtained by several Skyrme and different variants of relativistic mean-field models. The different Skyrme models used are SLy4 \cite{Chabanat1998}, SLy5 \cite{Chabanat1998}, SLy6 \cite{Chabanat1998}, SLy7 \cite{Chabanat1998}, SkO \cite{Reinhard1999}, SIII \cite{Beiner1975}, SkM \cite{Krivine1980}, SkP \cite{Dobaczewski1984}, S255 \cite{Agrawal2003}, SkI3 \cite{Reinhard1995}, HFB9 \cite{Goriely2005} and RMF models are NL3 \cite{Lalazissis1997}, NL3* \cite{Lalazissis2009}, FSU-Gold \cite{Todd-Rutel2005}, FSU-Garnet \cite{Chen2015}, DD-PC1 \cite{Niksic2008}, DD-PCX \cite{Yuksel2019}, DD-ME2\cite{Lalazissis2005}, DD-MEa to DD-MEe \cite{Vretenar2003}. This will give a crude idea about the sensitivity of the cross-section on the neutron skin thickness and the model dependence. However, the more rigorous calculation by fixing the proton radii within the various model need to be considered. We display in Fig. \ref{bar} the barrier heights and widths at a fixed energy. The widths of the barrier for a given nucleus are calculated at an energy equal to 0.7 times the maximum height of the barrier for SLy4 interaction. These barrier parameters are plotted in Fig. \ref{bar} for several Skyrme and relativistic mean-field models with PDD1 interaction as a function of neutron skin thickness. To guide our eyes we have connected the data with a straight line for Skyrme and relativistic mean-field models, separately. The variation of barrier parameters with the neutron skin for both the Skyrme and RMF models have similar trends but, the RMF models tend to yield smaller values. The overall decrease in the height, as well as the width of the barrier on increasing neutron skin thickness, is evident from the figure for both $^{54}$Ca+$^{54}$Ca and $^{124}$Sn+$^{124}$Sn. The influence of such variation of barrier parameters should also be reflected by the astrophysical $S$-factor for the considered nuclei. Fig. \ref{cs-s} elucidates the clear dependency of the $S$-factor on the neutron skin thickness. The $S$-factor readily changes by an order of magnitude for $^{54}$Ca and about two orders of magnitude for $^{124}$Sn with an increase in neutron skin thickness. A few points which are quite off from the trend are also found to have a larger deviation in binding energy in comparison to the other models.

\section{Conclusion}
We have performed the calculations of the cross-section for the sub-barrier fusion and astrophysical $S$-factor for several asymmetric nuclei. The double-folding model potentials required to compute cross-sections are obtained by folding nucleon-nucleon interactions with the density profiles for nucleons inside the nucleus obtained from mean-field models. The different nucleon-nucleon interactions considered are M3Y-Paris with and without density dependence and São Paulo potential version 2. The mean-filed models employed are based on the non-relativistic Skyrme-type effective interactions and different variants of relativistic effective Lagrangian associated with a wide range of the symmetry energy slope parameter or the neutron skin thickness. The barrier parameters such as its height and width decrease with an increase in neutron skin thickness which leads to the enhancement of cross-section and astrophysical $S$-factor up to one or two orders of magnitude. The sensitivity of the cross-section or the $S$-factor to the neutron skin thickness grows stronger with the increase in the proton number. The precise measurement of sub-barrier fusion cross-section or astrophysical $S$-factor in asymmetric nuclei may provide an alternative probe to determine the neutron skin thickness or the slope parameter of the symmetry energy.
 \section*{Acknowledgments}
We are thankful to B. Maheshwari,  M. Bhuyan, and Chandan Ghosh for the useful discussions. TG acknowledges the Council of Scientific and Industrial Research (CSIR), Government of India for fellowship Grant No. 09/489(0113)/2019-EMR-I. GS and BKA acknowledge partial support from the SERB, Department of Science and Technology, Government of India with grant no. SIR/2022/000566 and CRG/2021/000101, respectively.
\medskip
\typeout{}
%
\end{document}